# Applications of bound states in the continuum in photonics

Meng Kang[1], Tao Liu[2], C. T. Chan[1†], and Meng Xiao[2,3†]

[1]Department of Physics, The Hong Kong University of Science and Technology, Hong Kong, China

[2]Key Laboratory of Artificial Micro- and Nano-structures of Ministry of Education and School of Physics and Technology, Wuhan University, Wuhan, China

[3]Wuhan Institute of Quantum Technology, Wuhan, China

†email: phchan@ust.hk, phmxiao@whu.edu.cn

**Abstract**

Bound states in the continuum (BICs) have attracted attention in photonics owing to their interesting properties. For example, BICs can effectively confine light in a counter-intuitive way and the far-field radiation of photonic structures that exhibit BICs manifests fascinating topological characteristics. Early research into photonic BICs was primarily focused on designing artificial structures to produce BICs. However, since the mid-2010s, exploring the potential applications of BICs has been a growing trend in research. In this Review, we detail the unique properties of BICs, including the ability to achieve enhanced light confinement, sharp Fano resonances, and topological characteristics. We also explore phenomena derived from BICs including the generation of circularly polarized states and unidirectional guided resonances and the impact of BICs on various applications such as lasing, nonlinear frequency conversion, waveguiding, sensing and wavefront control. We also discuss the insights provided by BICs in several emerging research frontiers, such as parity–time symmetric systems, higher-order topology, exciton–photon coupling, and moiré superlattices.

**TOC Blurb**

Photonic systems provide a versatile platform to explore and use bound states in the continuum. This Review explores the potential of these states for enhancing light–matter interactions in various applications and investigating the physics of emerging photonic systems.

**Key points**

- Photonics provides a versatile platform to study and exploit the properties of bound states in the continuum (BICs), leading to a wide range of applications.
- The ability of BICs to achieve highly efficient light confinement leads to the coherent field enhancement of both electric and magnetic fields, which can improve lasing performance, nonlinear conversion efficiency, and waveguiding in photonic integrated circuits.
- Light scattered by photonic structures exhibiting BICs manifests the features of Fano resonances, leading to advanced functionalities in refractometric sensing and the identification of molecular fingerprints.
- BICs are characterized as topological polarization vortices in far-field radiation, in which the geometric phases in momentum space can be used to manipulate light, giving rise to polarization conversion, vortex beam generation, and beam shifts.
- BICs have been integrated to explore several emerging frontiers including parity–time symmetric systems, higher-order topology, exciton–photon coupling, and moiré superlattices.

**Introduction**

Light trapping has an indispensable role in light–matter interactions and paves the way for the development of various types of optical cavities[1,2]. Optical cavities can be created simply by assembling two reflectors to confine light between them. Progress in miniaturizing optical devices is driven by the ability to concentrate light in small volumes, which is enabled by innovations in the design of microcavities and nanocavities. An ideal cavity would infinitely confine light at the desired frequencies. If the frequency of a discrete optical mode is outside the frequency range of the continuous spectrum of the propagating waves, the light is trapped inside the cavity because there is no pathway for it to couple with the propagating waves. For instance, in distributed Bragg reflectors or photonic crystal defect cavities, light is confined by a photonic band gap, which prevents leakage to bulk modes in the surrounding structures. Resonators can also confine light; for example, whispering-gallery modes use total internal reflection at a dielectric interface to concentrate energy almost infinitely in microcavities such as microspheres, disks or ring waveguides[3,4]. However, when the frequency of the discrete mode falls within the frequency range of the continuous spectrum, the light inside the cavity couples with the propagating waves and leaks out. This radiation compromises the efficacy of light confinement in a cavity. The ability of a cavity to confine light is characterized by a finite quality factor, which is usually defined as $Q = \frac{\omega_0}{2\gamma}$, in which $\omega_0$ and $\gamma$ indicate the resonance frequency and decay rate of the cavity, respectively, with a high $Q$ value indicating that the cavity has a strong confining ability.

Bound states in the continuum (BICs) are trapped with an infinite lifetime despite being embedded in the continuum spectrum of extended states. In 1929 von Neumann and Wigner[5] proposed the existence of BICs, suggesting an artificial potential profile that could support BICs for matter waves. After being considered a mathematical curiosity for decades, in the 1960s and 1970s it was realized that BICs were universal wave phenomena of interfering resonances and could be found in both quantum and classical systems[6,7]. The flexibility of artificial photonic structures such as waveguide arrays and metasurfaces provides opportunities for realizing topological phenomena that would be difficult to demonstrate in their quantum counterparts. Various mechanisms for achieving BICs in photonic systems have been proposed and experimentally demonstrated[7]. Unlike conventional high-$Q$ cavities, BICs confine light in the continuum, and the volume of the confined mode can reach nanoscales with transparent dielectric materials. Research into BICs in photonic crystal slabs **[G]** (PCSs) offers a useful approach for achieving high-$Q$ cavities that is easier than using cavities in three-dimensional (3D) band gap materials and more robust than conventional approaches because BICs are topologically protected[8].

BICs are supported in systems that extend to infinity in at least one direction. Such systems have a finite number of radiation channels, and hence the radiation coefficients can be eliminated through symmetry mismatch with the modes in the continuum or by tuning a finite number of independent parameters to achieve destructive interference in the far field. In contrast, finite-sized compact structures have infinite numbers of radiation channels, which outnumber the parameters in the system that can be tuned to suppress far-field radiation, leading to the absence of true BICs in such systems[7,9]. Exceptions can occur when a material has extreme permittivity ($\varepsilon$) and permeability ($\mu$) values for example, $\varepsilon = \pm\infty$, $\mu = \pm\infty$, $\varepsilon = 0$ or $\mu = 0$, whereby all the radiation channels are tuned to vanish simultaneously. As theoretically proposed, nonradiative BIC modes can be trapped by compact meta-atoms if they are embedded in zero-index materials **[G]** [10-15].

The emergence of BICs can be interpreted through concepts of symmetry mismatch or destructive interference. Symmetry mismatch indicates that modes belonging to different symmetry classes are decoupled and symmetry operations classify eigenmodes into different representations. A discrete mode and the continuum modes cannot couple if they belong to different irreducible representations, leading to symmetry-protected BICs. For instance, any non-degenerate states at the center of Brillouin zone ($\Gamma$ point) of a PCS with a $C_{4v}$ point group **[G]** symmetry are BICs because they belong to different symmetry classes to the free propagating states in the light cone[16-20]. Destructive interference provides a fundamental mechanism to generate BICs, in which two or more radiation channels cancel one another in the far field. For instance, two resonances with the same mirror symmetry in a PCS with $C_{4v}$ symmetry can couple to form anti-crossing bands, resulting in a BIC near the anti-crossing point[21], known as a Friedrich–Wintgen BIC[6]. One single resonance can also evolve into BICs, if the single resonance is considered to be derived from the coupling of a set of coupled modes[22-25] or electromagnetic multipoles[26,27], allowing for the possibility of destructive interference. This mechanism can be used to explain scenarios in which the radiation at some specific wave vectors in a PCS was accidentally eliminated[28]. However, classifying the mechanism that leads to the formation of BICs as either symmetry mismatch or destructive interference is not always definitive. In certain instances, symmetry-protected BICs can also be described in the context of destructive interference[25].

In this Review, we present an overview of photonic BICs and their applications. There are already some useful reviews about BICs, which focus either on the mechanisms or specific applications of BICs such as nonlinear effects, lasing and sensing[7,9,29-34]. Here, we review the unusual properties of photonic BICs, provide insights on the impact of BICs in various applications, and envision the use of BICs in future research. We provide details and insights on why BICs are useful for some applications. For example, we first describe the enhanced light confinement induced by BICs and their ability to improve lasing performance, boost nonlinear conversion efficiency, and upgrade waveguiding in photonic integrated circuits. We then discuss Fano resonances with tailored ultra-narrow line widths and the implications for sensing and metasurfaces. We also analyze the topological characteristics of BICs and the emergence of polarization vortices around BICs, and their contributions to wave front control including the realization of polarization conversion, vortex beams, and beam shifts. Finally, we provide an outlook of the impact of BICs in several emerging frontiers, such as parity–time (PT) symmetric systems, higher-order topological phases, moiré superlattices and exciton–photon coupling.

**Enhanced light confinement**

BICs concentrate energy by eliminating radiation loss, facilitating light confinement. For instance, the mode amplitude of guided resonances of PCSs is primarily confined to the region near the slabs. In theory, BICs can trap light near the subwavelength-thick slabs with an infinite $Q$ factor. BICs with $Q$ factors of more than $10^6$ have been experimentally observed in PCSs[20,28]. Near-field mapping has been used to directly probe the localized light fields of BICs in metasurfaces[35]. Light localization has also been experimentally demonstrated by observing the propagation of defect states, wherein the defect states were BICs embedded in the extended states of one-dimensional (1D) horizontal coupled-waveguide arrays[36-40]. Anisotropy can also induce BICs, leading to light confinement in photonic systems such as anisotropic waveguides[41,42], 1D photonic crystals with an anisotropic defect layer[43,44] and materials with intrinsic anisotropy such as $LiNbO_3$. BICs can also be constructed to confine light in structures such as coupled-waveguide arrays with engineered hopping rates[45,46], photonic crystal fibers[47], nanowire geometric superlattices[48-50], stacked PCSs[51-55], the surface of PCSs[56], environment-engineered PCSs[57,58], PCSs with higher-order diffraction[59] and hybrid plasmonic-photonic gratings[60]. An infinite $Q$ factor is not attainable in practical systems owing to intrinsic absorption by the material and the inevitable scattering due to imperfections arising from the fabrication process. In addition, all real samples have finite sizes. Nevertheless, BIC-based ultrahigh-$Q$ resonances, usually called quasi-BICs, can confine sufficient energy in finite-size structures for them to be useful in various applications. For instance, in high-index dielectric nanoresonators, strong coupling between Mie resonances **[G]** and Fabry–Pérot resonances **[G]** can be tuned to suppress radiation and fulfill the quasi-BIC conditions[61]. Quasi-BICs can also be confined in micro-ring structures consisting of radially distributed rod elements[62], photonic crystal defect nanocavities with a line-defect mode[63] and truncated PCSs[64,65]. The total $Q$ factor of finite structures is determined by the radiative $Q$ factor $Q_{rad}$ and the dissipative $Q$ factor $Q_{dis}$, with high $Q_{rad}$ and $Q_{dis}$ values indicating small amounts of radiative and absorption loss, respectively. When the system reaches the critical coupling condition, $Q_{rad} = Q_{dis}$, the maximum field enhancement is achieved with an external incident light. Increasing $Q_{dis}$ by reducing material loss can further increase the field enhancement achieved at the critical coupling condition[66,67].

Light trapping by BICs enables coherent field enhancement, which is beneficial to light–matter interactions because the interaction strength and interaction time are both enhanced, improving the performance of many photonic devices. BICs and quasi-BICs have been used to reduce the lasing threshold[63,68-74], increase the efficiency of nonlinear frequency conversion **[G]** [67,75-84], and mitigate the radiation loss associated with waveguiding in photonic integrated circuits[85-97]. Moreover, owing to their ability to concentrate energy in nanostructures, (quasi-)BICs can be used to miniaturize optical devices[63,68,73,79].

**Lasing**

The ability of BICs to suppress radiation loss boosts light confinement, thereby opening possibilities to reduce the pumping threshold and size of lasers. Although it is not explicitly emphasized, the actions of many surface-emitting lasers can be interpreted as using symmetry-protected BICs, which have large-area coherent oscillations, facilitate control of the lasing pattern, and have a high-power output[98]. BICs are driving improvements in various aspects of lasing performance. For example, in an array of suspended cylindrical nanoresonators (**FIG. 1a**), the lasing threshold was reduced to a minimum of 56 μW at the BIC mode[68]. Such a BIC mode is robust and the lasing action persists even when the array is shrunk to a small array of 8-by-8 cylindrical nanoresonators. However, the finite size of the array inevitably broadens the modes in momentum space, which introduces radiative loss, limiting the extent to which the lasing cavity can be miniaturized.

To overcome this challenge, multiple off-Γ BICs can be simultaneously tuned to the vicinity of symmetry-protected BICs to suppress out-of-plane radiation over a broader wavevector range[99]. Based on this concept, ultralow-threshold lasers could be made within a small finite-size photonic crystal cavity[69]. BIC lasers can be further miniaturized[73] by combining them with laterally confined photonic bandgaps[65]. When light is confined within subwavelength dielectric nanoresonators, lasing action can occur at a quasi-BIC resonance, which is formed through the destructive interference between the Fabry–Pérot and Mie modes[71]. When lasers are miniaturized to the microscopic regime, quantum fluctuations become the dominating factor in limiting the line width of the laser. For these small lasers, BICs can improve the $Q$ factor of the cavity to effectively quench the quantum noise. A BIC mode can be formed by engineering the Fano interference **[G]** between the continuum in the waveguide and discrete modes in the nanocavity[63](**FIG. 1b**). Here the gain material is only inside the waveguide and the nanocavity in the line defect is passive. The field of the BIC mode is mostly concentrated in the passive nanocavity; thus the corresponding $Q$ factor (~ 78,000) is high and consequently the line width of this microscopic laser is substantially reduced so it is more than 20 times smaller than that of previously reported microscopic lasers.

In addition to enabling the miniaturization of lasers, BICs can also provide coherent field enhancement over a large area to increase the power output of surface-emitting lasers. However, the appearance of high-order transverse modes as the lateral size of the laser is increased has hindered progress in this direction. To tackle this scaling challenge, a mechanism based on zero-index modes using BICs was proposed to maintain single-mode lasing as the lateral size of the laser is increased[100]. A surface-emitting laser was fabricated with $C_{6v}$ point group **[G]** symmetry, in which the radius of the holes in the structure is finely tuned so that the non-degenerate mode of the $B_1$ irreducible representation and the doubly degenerate modes of the $E_2$ irreducible representation from the $C_{6v}$ point group, which are all BICs, become degenerate at the $\Gamma$ point (**FIG. 1c**). As a result, zero-index BICs are created through a Dirac-like cone dispersion at the center of the Brillouin zone. When such a system is finite in the lateral direction, the fundamental mode is a pure $B_1$ mode and all the other higher-order transverse modes are a hybridization of the $B_1$ and $E_2$ modes. In this system the laser cavity was truncated in the lateral direction such that it only supported the $B_1$ mode, as fundamental modes owing to the six-fold rotational symmetry and the $E_2$ mode with only two-fold rotational symmetry decayed more quickly than the $B_1$ mode. When the Dirac-like cone is achieved, the $E_2$ mode becomes dominant in the higher-order mode as the size of the laser is increased. Consequently, the contrast in $Q$ between the fundamental mode and all the other higher-order modes tends towards a constant value with the increasing size of the laser, making the fundamental mode have a competitive lasing threshold. In addition, because the fundamental mode effectively has a refractive index of zero[101] it locks the field of the whole sample in phase, leading to single-mode lasing. Besides zero-index BICs, BICs in a single band can also generate a size-invariant $Q$ contrast between the fundamental mode and higher-order modes[102], which is promising for achieving a scalable single-mode laser.

Using higher diffraction orders to partially break the BIC condition offers a way to control the direction of the emitted light and the $Q$ factor of the resulting quasi-BICs. The introduction of diffractive orders into arrays of dielectric nanoantennas (**FIG. 1d**), was used to partially break a BIC into a quasi-BIC whilst maintaining strong light confinement (large $Q$ factor)[70]. The direction of the laser emission was controlled through the leaky quasi-BIC channel from the higher-order diffraction by tunning the period of the nanoantenna array along one of the primitive axes. The introduction of controllable radiation channels through diffraction can also improve the external quantum efficiency of BIC lasing[74]. Using the polarization singularity of first-order diffraction as a high-$Q$ lasing mode, BIC lasing with high external quantum efficiency has been demonstrated by coupling first-order diffraction with highly efficient radiation channels at zero-order diffraction[74].

BICs can also be used to generate chiral emissions by constructing chiral quasi-BICs[103-106]. In optically thin structures such as PCSs or metasurfaces, lasers emitting purely left or right circularly polarized light remain elusive because chiral light–matter interactions are typically weak. Metasurfaces possessing a circularly polarized state (called a C point) at the $\Gamma$ point with a high $Q$ factor offer a promising solution. When the in-plane symmetry $C_2^z$ ($C_m^z$ indicates $\frac{2\pi}{m}$ rotation symmetry around the $z$ axis, and $m$ is an integer) of a metasurface is broken, a BIC breaks into two C points, owing to the need for topological charge to be conserved[107,108]. When up–down mirror symmetry $\sigma_h$ is also broken, one of the C points can be tuned to the $\Gamma$ point where the $Q$ factor remains high[106]. The combination of the large field enhancement at the $\Gamma$ point and the chirality of the C points provides an efficient scheme for chiral emission. With this strategy, chiral photoluminescence and lasing have been realized (**FIG. 1e**), exhibiting a high handedness purity, directionality, and $Q$ factor[72].

**Nonlinear frequency conversion**

The enhancement of BICs in the near field is also beneficial in nonlinear optics for improving the efficiency of nonlinear optical effects. Compared to their plasmonic counterparts[109,110], high-index dielectric nanostructures have low material losses and provide both electric and magnetic field enhancement in bulk volumes[111], leading to highly efficient nonlinear frequency conversion. However, the nonlinear efficiency of dielectric nanostructures remains limited by radiative loss. Quasi-BICs and BICs can provide a viable solution by suppressing this radiation. Subwavelength dielectric resonators have been used to empower high efficiency frequency conversion by exploiting the destructive interference between Mie resonances and Fabry–Pérot resonances in quasi-BICs[75,78,79,82,112]. For instance, the efficiency of the second harmonic generation (SHG) at a quasi-BIC resonance of an individual dielectric nanoresonator (**FIG. 2a**) is at least two orders of magnitude higher than that achieved with conventional approaches[79]. When a nanostructure is subjected to a strong enough excitation, nonlinear effects enter the nonperturbative regime, leading to high harmonic generation[113]. For example, quasi-BICs have been used to enhance high harmonic generation in a subwavelength dielectric resonator[112]. The yield of this high-harmonic generation follows a power scaling law that is distinct from perturbative scaling and exhibits saturation, owing to its nonperturbative origin.

Besides using nanoparticles, the nonlinear optical processes in resonant metasurfaces can also be substantially boosted by manipulating the linear optical response in the quasi-BIC region through structure optimization[67,76,80,83,84,114]. Phase-matching requirements are crucial constraints in bulk nonlinear crystals but they can be circumvented in metasurfaces as the nonlinear processes occur within a region of subwavelength thickness[84]. However, the limited thickness of nonlinear materials might reduce the nonlinear efficiency of the metasurfaces. The field enhancement from quasi-BICs has been used to achieve high efficiency harmonic generation in dielectric metasurfaces[67,76,80,83,114], including high harmonic generation with nonperturbative features such as the saturation in power scaling[83]. The $Q$ factors, the coupling efficiency and the resonance wavelength of metasurfaces can be engineered by converting symmetry-protected BICs to quasi-BICs through symmetry breaking[115]. The high $Q$ factor at quasi-BIC resonances can also be used to boost the emission of entangled photons from spontaneous parametric down-conversion in semiconductor metasurfaces[84] (**FIG. 2b**). These nonlinear metasurfaces can be engineered to host quasi-BICs at single or several wavelengths. Combining these structures with multiple pumping wavelengths enables the generation of complex quantum states.

**Waveguiding**

BICs are also effective in mitigating the radiation loss of waveguiding in photonic integrated circuits. Zero-index materials are particularly interesting in the field of light–matter interactions because of their fascinating properties such as wavefront tailoring and large coherent length[116,117]. Losses from material absorption and radiation in zero-index materials prevent them from being used extensively to guide optical waves. To circumvent intrinsic material absorption loss, dielectric photonic crystals have been designed to exhibit a vanishing refractive index by constructing a Dirac-like cone dispersion at the Γ point[101]. However, these Dirac-like cones exist inside the light cone; thus zero-index PCSs suffer from radiative losses. This problem could be addressed by the light confinement enabled by BICs. One possible approach is to construct hexagonal photonic crystals with a triple degenerate Dirac-like point with BICs. In this case, all the modes of the PCS at the Γ point belonging to 1D irreducible representations and a two-dimensional (2D) irreducible representation $E_2$ of the $C_{6v}$ point group are symmetry-protected BICs. By designing the shape of the meta-atoms in the PCS (**FIG. 2c**), the $B_1$ (or $B_2$) and $E_2$ representations can be tuned to form a triple accidental degeneracy, leading to an effective zero-index metamaterial in which the eigenmodes are BICs[85]. Similar ideas can also be applied to representations that do not support symmetry-protected BICs by instead using destructive interference to induce BICs[87]. By adjusting the structure parameters, off-Γ BICs can be tuned to the Γ point while maintaining a Dirac-like cone at the Γ point. Following this strategy, a nonradiative Dirac-like cone can be formed with both a monopole mode and a dipole mode[86] at the Γ point, which therefore has both a permittivity and permeability equal to zero according to the effective medium theory[118].

Light confinement in waveguides and cavities has a vital role in advancing the miniaturization of photonic integrated circuits and it typically requires the refractive index of the material to be much higher than that of the surroundings to function as potential wells to confine light and minimize power dissipation. However, these high-refractive-index materials are not necessarily compatible with state-of-the-art nanofabrication technologies, which hinders their integration in photonic integrated circuits. BICs provide a way to trap and guide light using low-refractive-index materials. One theoretical design proposed that depositing a low-refractive-index waveguide on a high-refractive-index membrane can eliminate optical dissipation owing to the BICs arising from the destructive interference of different dissipation channels[88]. Such photonic integrated circuits have since been developed by patterning low-refractive-index materials such as organic polymers rather than etching the high refractive index membrane such as single-crystal lithium niobate ($LiNbO_3$) [90] (**FIG. 2d**). Based on this architecture, various photonic components have been experimentally demonstrated, such as directional couplers[90], Mach–Zehnder interferometers[90], electro-optic modulators[90], multi-channel mode (de)multiplexers[91] and acousto-optic modulators[92]. Hybrid photonic devices have also been achieved by integrating 2D materials into these architectures[93]. In addition, high-efficiency SHG has also been demonstrated with this fabrication-friendly platform[94-96].

**BIC assisted sensing**

Fano resonances occur when a discrete bound state interacts with a continuum of bright states[119-121]. Ideal BICs are decoupled from the continuum spectrum, and cannot be probed through the optical response to external excitations. A small change in system parameters can turn BICs into quasi-BICs, which manifest as sharp Fano resonances in the light scattered from the system[28,115]. PCSs can be used to demonstrate Fano resonances that originate from BICs. Under illumination, guided resonances of PCSs (discrete modes) are excited and interfere with the broader Fabry–Pérot resonances (continuum of states) to produce Fano resonances in the transmission and reflection spectra[19]. The amplitude of a Fano resonance varies sharply with a line width of $\frac{\omega_0}{Q}$. Close to a BIC in momentum space this line width $\frac{\omega_0}{Q}$ decreases and vanishes at the BIC[20,28]. Besides PCSs, Fano resonances also appear in the light scattered by high-index dielectric resonators, in which Mie resonances and Fabry–Pérot resonances couple to form quasi-BICs[61,122,123]. In side-coupled waveguide arrays, Fano resonances form as a result of the interference between different decay channels, and destructive interference between these channels gives rise to BICs[37,124-127]. Fano resonances derived from anisotropy-induced BICs have been observed in 1D photonic crystals with a defect anisotropic layer[43,44]. Despite being constrained by intrinsic material loss, plasmonic BICs can still produce strong near-field enhancement and relatively sharp Fano resonances[60,128-131].

Metasurfaces provide a versatile platform for tailoring Fano resonances derived from quasi-BICs. For example, a metasurface consisting of a square array of tilted nanobar pairs[115] (**FIG. 3a**) possesses a symmetry-protected BIC when the tilted angle $\alpha$ is zero. When $\alpha$ becomes nonzero the symmetry-protected BICs evolve into quasi-BICs with finite $Q$ factors. The transmission spectra from this structure exhibit Fano resonances with asymmetric line shapes at quasi-BICs (**FIG. 3b**). The Fano asymmetric parameter, which describes the asymmetricity of the line shape, can be tuned to diverge in the vicinity of quasi-BICs by varying the structural parameters[61,123,132], meaning that the spectrum becomes symmetric. Tuning the structure parameters to approach the symmetry-protected BIC at $\alpha = 0°$, gradually sharpens the Fano resonances until they eventually vanish. Thus, both the transmission line width and asymmetric parameter of Fano resonances can be tailored in the vicinity of BICs, paving the way for future applications such as nonlinear optical effects and sensing.

Optical sensors based on (quasi-)BICs offer exciting prospects for developing ultra-sensitive chemical and biological detection. Sensors with dielectric nanostructures that exploit (quasi-)BICs can mitigate the problem of intrinsic absorption loss associated with their plasmonic counterparts[133,134]. Additionally, dielectric structures are more compatible with the standard complementary metal-oxide semiconductor (CMOS) fabrication processes than plasmonic structures, facilitating nanoscale integration and industrial manufacturing. Dielectric nano-resonators with quasi-BICs also offer good sensing performance because they provide ultra-sharp resonances and substantial near-field enhancement for both electric and magnetic fields. These advantages of quasi-BICs have enabled the development of advanced sensors with exceptional performance in refractometric sensing[135-143], surface-enhanced spectroscopy[144-147], and chiral sensing[148]. For instance, by combining quasi-BIC dielectric metasurfaces with hyperspectral imaging and data science techniques, an ultrasensitive label-free sensing platform has been developed that does not require the use of spectrometers[139]. By varying the geometric parameters of the metasurface structure for example by scaling the lateral dimensions of the tilted nanobar array introduced above (**FIG. 3a, bottom**), the resonance frequency can be tuned continually while keeping the $Q$ factor high. The sharp resonance provides high spectral resolution and the high-$Q$ value ensures large near-field enhancement. Different sensors, each tuned to a different quasi-BIC resonance frequency can be assembled into an array (**FIG. 3c**). CMOS pixels below each sensor measure the spectral response of the sensor with and without analytes. This quasi-BIC based sensing scheme achieved a detection limit below 3 molecules $\mu m^{-2}$. Furthermore, because this sensor array reads out the molecular absorption signatures at different quasi-BIC resonance frequencies (**FIG. 3d**) the protein fingerprint can also be mapped with high sensitivity[144]. This approach yields barcode-like spatial absorption maps for imaging and can perform chemical identification and composition analysis on surface-captured analytes. In addition, owing to the dependence of the quasi-BIC resonance frequency on the incidence angle, angle-resolved measurements can be used to detect molecular absorption fingerprints over a broad spectrum without the use of a tunable source[146].

In addition to high $Q$ factors and field enhancement, increasing the coupling efficiency between light and the sensors is also important to achieve the sensitive enhancement of molecular absorption spectra. Tailoring the radiative decay rate of quasi-BICs in plasmonic metasurfaces to match the rate of intrinsic loss revealed that the performance of surface-enhanced molecular sensing strongly depends on whether the coupling region is in the under-coupling, critical-coupling, or over-coupling regime[147]. The sensing enhancement that can be achieved with quasi-BIC resonances can also be applied to circular dichroism (CD), which typically involves weak signals. Quasi-BICs can generate strong superchiral fields to enhance CD sensing. For instance, the coalescence of transverse electric (TE) and transverse magnetic (TM) quasi-BICs can generate strong superchiral fields[149]. It has been suggested that refractometric sensing and CD spectroscopy could be combined to enable the analysis of enantiomer composition[148]. Surprisingly, the weak chirality of enantiomers could induce strong coupling between two quasi-BIC resonances in a metasurface, in which the hybrid modes represent the substantially enhanced CD signals[150].

**Exploiting topological robustness**

**Topological origin of BICs**

For BICs to be used in various applications it is important that their existence does not depend on extreme precision in system parameters. Indeed, BICs are robust against parameter variation in many optical systems such as PCSs. Generally, when the number of tunable parameters is equal to the number of equations required to satisfy the BIC condition, the presence of the BIC is robust against local perturbation. Note that each complex equation is counted as two real equations. The robustness of BICs can be understood with a topological argument. The BIC can be defined as the origin point $\{0\}$ in the Euclidean space $\mathbb{R}^n$ of dimension $n$ spanned by $n$ equations that are required to satisfy the BIC condition. After removing the origin, $\mathbb{R}^n - \{0\}$ is homotopically equivalent to the $(n-1)$-sphere $S^{n-1}$, that is, $\mathbb{R}^n - \{0\} \simeq S^{n-1}$. The ($n$–1)th homotopy group $\pi_{n-1}(S^{n-1})$ of $S^{n-1}$ is isomorphic to the integer group $\mathbb{Z}$, that is, $\pi_{n-1}(S^{n-1}) = \mathbb{Z}$, where $\pi_{n-1}(S^{n-1})$ classifies the mapping from one $S^{n-1}$ (parameter space) to another $S^{n-1}$ (space of equations to be satisfied). Consequently, a BIC can be characterized by an integer topological invariant. For example, consider the case when the number of parameters and the number of equations are both equal to two. In this case both the parameters and equations form 2D hyper-surfaces, which are denoted as the parameter space and target space, respectively. A loop enclosing the BICs in the parameter space can be continuously mapped to a closed loop in the target space. The fundamental group that characterizes this mapping is the integer group, that is, $\pi_1(S^1) = \mathbb{Z}$. This integer labels the charge of the BICs. This argument is general and can be extended to higher dimensional spaces. For example, in 3D space, the loop becomes a sphere and the fundamental group is $\pi_2(S^2) = \mathbb{Z}$. As in the 2D case, the space formed by those three equations can be used to define the topological invariant in 3D.

The topology of BICs has been discussed under various contexts[8,151-153]. For instance, in an array of dielectric spheres, BICs are associated with the phase vortex of the quasimode coupling strength[151]. Another more common topological interpretation involves the BICs of guided resonances in PCSs. For a PCS with time-reversal symmetry (no magnetic response, which holds in most optical materials), and $C_2^z$ and up–down mirror symmetry $\sigma_h$, the far-field radiation can be described by a two-component polarization vector $\mathbf{E} = (E_x, E_y)$ in momentum space in which BICs present as polarization vortices[8,154,155]. Away from the $\Gamma$ point $\mathbf{E}$ is elliptically polarized and the polarization direction is the major axis direction of the polarization ellipse[156]. Here the emergence of a BIC requires both $E_x(k_x, k_y) = 0$ and $E_y(k_x, k_y) = 0$ in momentum space of $\mathbf{k} = (k_x, k_y)$, and the polarization vector $(E_x, E_y)$ defines the space $\mathbb{R}^2$ (**Box 1**). As a polarization singularity, the BIC carries a topological charge ($q$) that is defined more explicitly as the winding number of the polarization direction ($\theta$), that is, $q = \frac{1}{2\pi}\oint_C \nabla_k \theta(\mathbf{k}) \cdot d\mathbf{k}$, along an anticlockwise loop $C$ enclosing the BIC (**Box 1**). Topological charge can explain the existence and evolution of BICs because the total topological charge must be conserved during the generation, evolution and annihilation of BICs. Moreover, the distribution of BICs in momentum space is also related to the space group symmetry of the PCSs.

**Manipulating topological charges**

Understanding and using the rules of topological charge conversion can lead to many applications of BICs. For example, the topologically protected tunability of topological charges in momentum space enables BICs to be merged[99]. The structure parameters of a nanostructure can be varied to control the location of multiple BICs, each with a topological charge, and move them closer together in momentum space so they merge and function as a single BIC, known as the merging BIC [99] (**FIG. 4a**). Consequently, the $Q$ factors of radiative states near the merging BIC are substantially increased by more than one order of magnitude over a broad range of wavevectors (**Fig. 4a, middle),** which is essential for practical applications because scattering losses caused by fabrication imperfections can allow BICs to couple to nearby radiative states. Increasing the $Q$ factors of the nearby states makes merging BICs (**Fig. 4a,** **middle)** much more robust against scattering losses than the isolated BIC (**Fig. 4a, right**). Merging BICs at the Γ point can be assembled with off-Γ BICs that are related by spatial symmetries, whereas the formation of merging BICs with steerable momentum requires BICs that originate from alternative physical mechanisms[21,157].

Symmetry breaking perturbation offers a useful tool for manipulating topological charges. BICs with higher topological charges can exist at high symmetry points in the reciprocal space if the system has a high rotation symmetry. If this high rotation symmetry is broken but the $C_2^z$ symmetry is preserved, higher-charged BICs split into fundamental BICs with charge $q = \pm 1$. For example, a higher-charged BIC with $q = -2$ is split into two off-Γ BICs when the $C_6^z$ rotation symmetry is broken but the $C_2^z$ symmetry is preserved[158] (**FIG. 4b**). When either the $C_2^z$ or $\sigma_h$ symmetry is broken, BICs are no longer stable. In particular, when the $C_2^z$ symmetry is broken, BICs are destroyed and the integer topological charge splits into pairs of half topological charges[107] (**Box 1**), corresponding to C points with opposite chirality (left and right circularly polarized, **FIG. 4c**). As these two C points lie at the opposite poles of the Poincaré sphere, such a symmetry reduction process covers most of the polarization states. Conversely, two C points with the same half topological charge and opposite chirality can merge and form a new BIC[159]. Moreover, pairs of C points can be generated from the void[160,161]; in this case, the two C points carry opposite half charges and the same chirality. When the $\sigma_h$ symmetry is also broken, the far-field radiation is different in the regions above and below the nanostructure. On one side of the nanostructure the far-field radiation can be canceled selectively by merging two C points with identical charges but opposite chirality[108,160,161] (**FIG. 4d**). Whereas on the other side the far-field radiation remains as the two C points are separated from each other. In other words, unidirectional guided resonances can be achieved that only radiate on one side of the nanostructure.

Chiral quasi-BICs[103-106] supported on PCSs can be realized through symmetry breaking. When $C_2^z$ is broken, off-Γ C points are derived from a BIC and exhibit extrinsic chirality under oblique incidence[162-164]. Alternatively, to achieve intrinsic chirality, $\sigma_h$ symmetry perturbations must be introduced to tune a C point back to the Γ point[106] (**FIG. 4e**). Furthermore, the mode at the Γ point maintains a large $Q$ factor; therefore, it is a quasi-BIC. Thus, chiral quasi-BICs that exhibit strong chirality and high $Q$ factors can be achieved, which are useful in enhancing chiroptical effects in chiral light–matter interactions[72,165].

**Polarization vortices**

The nonzero topological charge of a BIC leads to a nonzero winding of the polarization states and hence a polarization vortex emerges from the BIC. Polarization vortices bring a new degree of freedom for engineering the optical response of a nanostructure in momentum space. In particular, the optical response of the system varies according to the polarization distribution associated with the topological charges in momentum space. Thus, the evolution of topological charges in momentum space when the structural parameters are varied and the symmetry of the system is reduced provides flexibility for light manipulation. In addition, controllable properties of incident light (such as the incidence direction, polarization, phase, wavelength, and light intensity) offer many further degrees of freedom for tuning the optical response on demand to achieve specific applications.

The polarization vortices associated with BICs can also be used to shape wavefronts in real space[166-172]. The variation of polarization states is inevitably accompanied by additional geometric phases (Pancharatnam–Berry phase)[173-175]. The evolution of polarization states on the Poincaré sphere provides a visual interpretation of the emergence of geometric phases (**Box 1**). The ability to engineer the geometric phase distribution in real space is needed to achieve wavefront shaping in momentum space, and much progress has been made in flat optics and structured light[176-178], including BIC-based nonlocal metasurfaces[104,179-183]. An alternative approach manipulates the geometric phase distribution in momentum space instead of real space by varying the polarization of the states rapidly around the polarization vortices as a function of the wavevector. This polarization variation gives rise to geometric phase gradients in momentum space, which in turn can generate vortex beams and shift the real space position of a light beam.

So far, polarization vortices have been used to achieve various applications, including polarization conversion[184-186], vortex beam generation[166,167,169], optical switches[168], and beam shifts[170-172].

*Polarization conversion*

The polarization vortices encircling BICs provide diverse polarization states, which enable the development of PCSs with large degrees of freedom in the manipulation of their optical responses. Polarization conversion is conventionally based on accumulated phase retardation, which usually demands bulky optical components. Metasurfaces to enable polarization transformation have been developed by designing the shape and orientation of the anisotropic elements in real space[187,188], which could facilitate the miniaturization and integration of optical devices. Guided resonances supported by PCSs could provide an alternative platform for polarization conversion in momentum space. These guided resonances interact with the incident light through their radiation channels; thus the optical response of the PCS can be modulated by changing the guided resonances. For example, the relative phase delay between different incident channels, the radiative coupling coefficients and the frequency of the input beams can be used to tune the system to achieve coherent arbitrary polarization conversion in the vicinity of a BIC[186].

The optical response of a PCS can be mathematically characterized by a scattering matrix, which establishes the relationship between the incoming and outgoing light following the interaction with the guided resonances. This scattering matrix can be illustrated by the polarization conversion achieved using a PCS structure with a periodic array of holes on top of a mirror[184] (**FIG. 5a, left**). In this case, the scattering matrix is a $2 \times 2$ complex matrix because the frequency of the incoming light is below the diffraction limit and the transmission channel is blocked by the mirror. The diagonal and off-diagonal terms of the scattering matrix represent the co-polarization and cross-polarization coefficients, respectively. The complete polarization conversion point appears at the topological vortex center of the complex co-polarization field in the momentum space. In addition, as the wavevector of the incident light is varied around this complete polarization conversion point, almost arbitrary polarization conversion can be achieved with output polarization states covering the whole Poincaré sphere[185]. The scattering matrix varies with frequency; therefore, the complete polarization conversion point should extend to a line spanning a broad range of frequencies owing to its nontrivial winding number. This complete polarization conversion line can also be understood with the critical coupling picture in which the coupling coefficients of the guided resonances with orthogonal polarizations (s and p) have the same magnitude. Because the coupling coefficients of both polarizations are zero (and hence equal) for a BIC, the BICs must be linked together by the complete polarization conversion line (**FIG. 5a**, right).

*Vortex beams*

BICs can induce a phase singularity in momentum space through polarization vortices, leading to the production of optical vortex beams. Optical vortex beams, which carry orbital angular momentum (OAM), have attracted widespread attention owing to their applications in micromanipulation, imaging, sensing and optical communications[189]. The cross-section of a vortex beam exhibits a phase distribution $\exp(il\varphi)$, where $l$ is an integer value indicating a quantized OAM of $l\hbar$ per photon. The positive and negative sign of $l$ is referred to as the left- and right-handed wavefronts, respectively, and $\varphi$ is the azimuthal phase. Traditionally, the generation of vortex beams requires bulk optical elements such as spiral phase plates and pitch-fork holograms, which has hindered attempts to miniaturize and integrate vortex beams. Progress in miniaturization has been made by exploiting microscale and nanoscale structures[190]. For example, metasurfaces provide platforms for generating vortex beams by using subwavelength scatterers to engineer customized spatial phase profiles[191-193]. However, these optical structures usually have optical centers and misalignment between the optical center and the center of the light beam can reduce the purity of the generated vortex modes and hence undermine the performance. Polarization vortices associated with BICs provide a convenient way to circumvent this challenge by tailoring the phase distribution of the wavefront in momentum space to achieve vortex beams using periodic structures. When light propagates through a PCS, the anisotropy of the far-field polarization leads to different scattering coefficients along the major ($t_a$) and minor axes ($t_b$) of polarization. For a guided resonance with a radiative polarization direction denoted by $\theta$, the scattered field can be derived from the Jones matrix ($\boldsymbol{T}$) in the right circularly polarized–left circularly polarized basis[166]:

$$\boldsymbol{T} = \frac{1}{2}\begin{pmatrix} t_a + t_b & (t_a - t_b)\mathrm{e}^{-2\mathrm{i}\theta} \\ (t_a - t_b)\mathrm{e}^{2\mathrm{i}\theta} & t_a + t_b \end{pmatrix}. \quad (1)$$

Here, $t_a$, $t_b$ and $\theta$ all depend on the wavevector. Thus, when the incident light is circularly polarized, the cross-polarized light scattered by a PCS acquires a Pancharatnam–Berry phase shift of $2\theta\sigma$, where $\sigma = -1$ if the incident light is right handed and $\sigma = +1$ if the incident light is left handed. Therefore, a spiral phase is acquired by the cross-polarized light owing to the polarization vortices around BICs where $\theta$ exhibits a nontrivial winding[166] (**FIG. 5b**). Highly efficient vortex beam generation can be achieved by blocking the transmission mode and increasing the ratio between the rate of radiative loss and the rate of intrinsic material loss[194,195] of the guided resonances.

Nonlinear geometric phases arising from nonlinear materials provide another way to tune the OAM of vortex beams[109]. For a circularly polarized fundamental wave, the nonlinear geometric phase of the *n*th harmonic beam is $(n \pm 1)\theta\sigma$, where $\theta$ is the polarization direction of states, $\sigma = \pm 1$ depends on whether the incident light is right handed or left handed, and the $\pm$ sign determines whether the chirality of the generated harmonic is the same or opposite to that of the fundamental wave. The gradient of the nonlinear geometric phase around a BIC can be used to obtain vortex beams carrying different OAMs from the harmonics generated by the incident waves[167].

*Vortex lasing*

(Quasi-)BICs with an ultra-high $Q$ factor facilitate ultra-low threshold vortex lasing. Symmetry-protected BICs become radiative quasi-BICs under symmetry breaking perturbations, leading to the distortion of the polarization vortices. This tunable property between BICs and quasi-BICs can empower optical switching between vortex lasing and linearly polarized lasing in an active PCS. For example, ultrafast all-optical switching was demonstrated in a perovskite-based vortex microlaser at room temperature (**FIG. 5c**), by controlling the spatial intensity profile of the pumping beams[168]. In addition to symmetry-protected BICs that enable lasing at the $\Gamma$ point, off-$\Gamma$ BICs can be used to achieve steering directional vortex lasing[169] because they can be tuned in momentum space owing to their topological protection.

*Beam shift*

BICs can be used to tailor the phase gradient of a wavefront in momentum space and therefore can facilitate enhanced beam shifts. Beam shifts are universal phenomena at the interface of inhomogeneous materials. There are two typical types of beam shifts: Goos–Hänchen shifts **[G]** stem from the dispersion of the reflection or transmission coefficients, whereas Imbert–Fedorov shifts **[G]** , also known as the spin Hall effect, are associated with the spin–orbital interaction of light[196-198]. The interaction and mutual conversion between spin angular momentum and OAM (both intrinsic and extrinsic) are manifestations of the spin–orbital interaction of light and thus naturally introduce beam shift. Geometric phases, stemming from the spin-redirection phase and Pancharatnam–Berry phase, underpin the polarization-dependent beam shifts. Phase gradients in real space lead to wave vector shifts in momentum space, whereas phase gradients in momentum space give rise to beam shifts in real space, manifesting a reciprocal relationship. Lateral beam shifts are usually restricted to subwavelength scales; therefore, attempts have been made to overcome this challenge[199,200].

By engineering quasi-BICs to achieve high-$Q$ resonances, sharp phase gradients in the reflection or transmission coefficients in momentum space become feasible[170], which can enhance the Goos–Hänchen shift by more than four orders of magnitude of wavelength. Additionally, the manipulation of polarization vortices has been explored as another approach to boost beam shifts[171]. Breaking the in-plane inversion symmetry of a PCS modifies the polarization distribution by splitting a symmetry-protected BIC into two C points with opposite handedness (**FIG. 4c**). If a linearly polarized ($\pm 45°$ as optimized) beam is applied to such a PCS at normal incidence, different geometric phases are imposed on the different wave vector components of the beam when it is cross-polarization-analyzed after passing through the slab. It has been demonstrated that the cross-polarization-converted beams can then be shifted by about 5 times the wavelength in real space owing to the momentum-space phase gradient between the two C points (**Fig. 5d**). Additionally, if a circularly polarized incident beam is used instead, the spin-Hall effect can be induced by the polarization vortices in the vicinity of BICs. This spin-Hall effect introduces a phase gradient in the cross-polarized state owing to the Pancharatnam–Berry phase and the additional resonant phase difference between guided resonances, leading to a spin-dependent beam shift[172].

**Outlook**

BICs have improved the performance of a wide range of applications that can benefit from strong light–matter interactions. Future work will continue to explore the rich physics of BICs in various photonic systems. Here, we envision several potential avenues for research in the near future.

BICs can be used to enhance lasing performance and promote miniaturization through optical pumping. Additional efforts could be devoted to harnessing BICs to improve the lasing performance of electrically driven lasers in which unidirectional guided resonances are favored. BICs can enhance nonlinear effects in both perturbative and nonperturbative regions, providing an excellent platform for exploring the underlying physics of frequency conversion, and its potential applications, such as high harmonic generation[83,112], optical-frequency combs[201,202], and the manipulation of quantum states[84,203]. The polarization of quasi-BICs can be manipulated to design nonlocal metasurfaces[104,179,181] and enhance chiral light–matter interactions[72,106,165]; research into such applications is still in a nascent stage. The active control of BICs is also necessary to achieve real-time and reversible manipulation of light[180,204,205]. Although the manipulation of BICs has primarily relied on spatial symmetry breaking, the breaking of Lorentz reciprocity can induce interesting effects such as nonreciprocal transmission[206], topological phase singularity pairs[153] and extended states in a localized continuum[207]. BICs can be further leveraged to improve the performance and miniaturization of on-chip devices, such as on-chip light sources, diffraction-free beam guiding[89], color pixels[208], and sensors[141,142]. Another exciting possibility is the combination of BICs with other emergent research fields such as PT symmetric systems, higher-order topology, exciton–photon coupling, and moiré superlattices, which will enrich findings from both sides. The following sections outline some examples of these potential applications of BICs.

**Non-Hermitian systems**

Non-Hermitian components bring unique properties to photonics such as complex spectra and exceptional points, which could facilitate the investigation of BIC-related topics. Photonic systems are popular platforms for exploring non-Hermitian systems, particularly the physics of exceptional points[209-211]. When the amount of gain and loss in photonic systems are balanced by varying a non-Hermiticity parameter, non-Hermitian systems obey PT symmetry and the corresponding Hamiltonian can exhibit entirely real eigenvalues[212]. However, when this balance between gain and loss is broken, the Hamiltonian undergoes a real-to-complex spectral phase transition at the exceptional points[213-215], at which both the eigenvalues and eigenvectors coalesce. In photonic systems loss can come from intrinsic material loss and radiative loss. Paired exceptional points can be produced by the radiative loss of PCSs. The non-Hermitian topological properties of these pairs of exceptional points lead to the formation of a bulk Fermi arc **[G]** and the far-field polarization of this system exhibits half-integer topological charges[216]. The topological charge of the bulk Fermi arc varies with the interplay between the paired exceptional points and other topological singularities such as C points and unidirectional guided resonances, which can be generated from BICs[161,217]. In particular, when C points[217] or unidirectional guided resonances[161] traverse a bulk Fermi arc and switch from one band to the other on Riemann surfaces, their topological charge remains invariant whereas the topological charge of the bulk Fermi arc changes to maintain overall topological charge conservation on each band (**FIG. 6a**). The change in the topological charge of the bulk Fermi arc alters the polarization distribution around it, which can lead to applications related to polarization manipulation. Additionally, investigations into photonic BICs and PT-symmetric non-Hermitian systems have been combined. For instance, in the continuum spectrum of a PT-symmetric optical lattice, the localization and power emission characteristics of the BICs are distinct when PT symmetry is maintained or broken[218,219]. The introduction of perturbations that do not break the PT-symmetry could split BIC modes, leading to the simultaneous emergence of lasing threshold modes and a new type of BIC, known as, PT-BICs[220,221]. Similar to conventional BICs, PT-BICs do not radiate; however, PT-BICs can be excited by external waves. The coexistence of BICs and exceptional points has also been demonstrated in PT-symmetric systems and anti-PT-symmetric systems[222-225], which suggests that unconventional BICs with the ability to achieve slow light and maximal coherent superposition are generated.

**Higher-order topological phases**

BICs in photonics open avenues for the investigation of higher-order topological phases. Higher-order topological phases exhibit a higher-order bulk-boundary correspondence such that a $D$ dimensional nontrivial bulk guarantees $D - d$ dimensional ($d = 2, 3, \cdots, D$) boundary modes such as corner or hinge states[226]. Photonic systems provide an important and flexible platform for designing and demonstrating higher-order topological phases[227]. Traditionally, the generation of higher-order topological states has mainly required confinement within a bulk band gap[226]. However, in 2019 it was shown that topological corner states can remain localized as BICs inside the bulk band region[228,229], which goes beyond the scope of traditional higher-order bulk-boundary correspondence. A coupled-waveguide array arranged in a second-order topological lattice with $C_{4v}$ and chiral symmetries was used to demonstrate that the corner states are symmetry-protected BICs and do not leak into the surrounding bulk states[230,231] (**FIG. 6b**). Additionally, higher-order topological BICs can be actively controlled by nonlinearity[232].

Another approach for realizing higher-order topological phases that exhibit BICs is to construct independent subspaces with different symmetries[233,234]. For instance, in mirror-stacked bilayer systems, topological corner states generated in one subspace decouple from the bulk states in another subspace owing to their different up–down mirror parities, thereby forming higher-order topological BICs[234]. A quadrupole topological insulator that exhibits corner states has been demonstrated in a photonic crystal that breaks time-reversal symmetry[235,236]. Owing to the absence of chiral symmetry in such systems, these corner states can be tuned into bulk bands to form topological BICs, as experimentally demonstrated in a gyromagnetic photonic crystal[236]. Compared with higher-order corner or hinge modes confined within a bulk bandgap, the presence of a bulk continuum offers more possibilities for exciting, identifying and using these higher-order topological phases. Higher-order topological BICs are important because photonic topological corner or hinge modes do not leak into the bulk states of a topological material but they can still couple with the modes inside the light cone and BICs are essential for providing tunable high-$Q$ surface states to suppress the radiation of topological corner or hinge modes.

**Moiré superlattices**

Moiré patterns emerge as a new degree of freedom for manipulating the optical response of photonic structures, providing another platform for studying BICs. Electronic moiré superlattices, based on twisted or lattice-mismatched van der Waals heterostructures, present a new class of electronic material[237,238] with interesting properties such as flat bands, correlated states and moiré excitons, which have inspired investigations into the use of moiré engineering in photonics. The ability to flexibly modulate the band structures and scattering properties has been explored in moiré superlattices comprising PCSs[239-244], in which the twist angle and the interlayer separation can be used to tune the interlayer coupling. Analogous to magic-angle graphene superlattices, photonic flat bands appear at certain magic angles in these moiré superlattices, which can induce light localization and increase the light–matter interaction[245]. The realization of BICs in moiré superlattices has been explored by twisting a bilayer PCS to produce a moiré superlattice, leading to the formation of photonic flat bands (**FIG. 6c**) [246]. The appearance of these photonic flat bands with narrow dispersion and high density of states improved the ability of quasi-BICs to increase light–matter interactions. In this bilayer structure the size of the moiré supercell is larger than the wavelength of the incident light; therefore, higher-order diffraction channels appear. The radiation from the zero-order diffraction channel vanishes when resonances at the Γ point have symmetry mismatch with the free propagating wave whereas the higher-order diffractions radiate lightly, and hence the bilayer PCS exhibits a quasi-BIC. When the twist angle is reduced, the total radiation from all of the diffraction channels is substantially suppressed and the $Q$ factor tends to infinity. Theoretical calculations have also suggested that these moiré quasi-BICs could be used to enhance the efficiency of SHG[246].

Many intriguing phenomena involving moiré quasi-BICs remain to be discovered. For instance, at certain angles, moiré superlattices become quasicrystals[237], which have higher rotation symmetry and can host BICs with higher topological charges[247]. Moiré patterns also produce a periodic potential profile; therefore, they provide a powerful approach for engineering excitons in transition metal dichalcogenides (TMDCs)[248]. By combining this approach with twisted optical cavities that support (quasi-)BICs, light–matter interactions can be greatly enhanced, making it easier to control excitonic behavior by modifying cavities.

**Exciton–photon coupling**

The use of BICs in exciton–photon coupling systems could make it possible to manipulate excitons and investigate phenomena such as exciton emission, nonlinear optical effects, and polariton BICs, in 2D materials including TMDCs. In weak coupling regions, BICs can increase the emission rate of excitons owing to the enhanced Purcell factor $F_p \propto \frac{Q}{V}$ ($V$ is mode volume) of the BIC[249]. In TMDC monolayers, excitons have large binding energies and dominate the optical properties of the monolayer at both cryogenic and room temperatures[250]. Excitons follow the spin-valley locking imposed by the optical selection rules; therefore, the direct interband transitions at the K and K′ valleys exclusively couple to right or left circularly polarized light, respectively. Through the chirality of C points, chiral quasi-BICs have been used to control the direction of valley exciton emission[251]. Optically forbidden dark excitons in TMDC monolayers originate from spin-forbidden optical transitions and have no in-plane dipole momentum, making them difficult to excite and detect with conventional far-field optical techniques. To probe dark excitons, strong in-plane magnetic fields or enhanced out-of-plane polarized light are required[252-255]. Transverse magnetic (quasi-)BICs can brighten dark excitons and tune their directional emission at room temperature[249]. For example, the photoluminescence of dark excitons in a PCS with a WSe$_2$ monolayer was substantially enhanced (by a factor of ~1,400) by at-Γ BICs and a highly directional emission was achieved owing to the coupling between the excitons and the off-Γ BICs. Integrating photonic systems possessing BICs with TMDC monolayers and heterostructures[248,256,257] enables light to be effectively confined; therefore, these integrated systems provide an exciting platform for improving the performance of exciton devices, such as exciton lasers[258-260]. 2D layered materials such as TMDCs and gallium monochalcogenides exhibit large nonlinear susceptibility; however, the nonlinear frequency conversion efficiency is limited by the atomic length of the light–matter interaction, owing to the thin nature of 2D layered materials[261,262]. To solve this problem, the field enhancement of dielectric metasurfaces at quasi-BICs has been used to substantially enhance nonlinear effects, such as SHG[261,262].

Exciton–polaritons are hybrid quasiparticles that stem from the strong coupling of cavity resonances and excitons. In the strong coupling regions, excitons of TMDC monolayers couple with BIC modes of the photonic cavity to generate polariton BICs[263-268]. Typically, TMDC monolayers are only deposited on the surface of a structure away from giant field enhancement of BICs, which is mainly localized within a bulk volume, preventing strong coupling being achieved at room temperature. To tackle this challenge, BICs originating from Bloch surface states[269] have been developed to promote surface field enhancement, resulting in record Rabi splitting and strong polariton nonlinearities at room temperature[268]. In addition, the high refractive index of TMDCs allows them to be structured as nanodisks[270] and metasurfaces that support BICs[271], enabling the generation of polariton BICs[272]. Owing to their hybrid nature, polariton BICs have long lifetimes and substantial nonlinearities inherited from their photonic and excitonic composition, respectively. These features have been experimentally confirmed by observed reflection spectra, photoluminescence spectra and polariton blueshifts[265,268]. Polariton BICs also carry topological features from their electromagnetic component such as polarization vortices of far-field radiation around BICs, which enables the polariton BIC to be observed and manipulated by incident light. For instance, polariton BICs have been achieved at room temperature in perovskite metasurfaces. The inherited topological properties of the polariton BICs in this structure were verified by the observation of polarization vortices[273,274]. Polariton BICs have also been used to promote polariton Bose–Einstein condensation (BEC), resulting in an extremely low threshold density for condensation[275]. For example, polariton BEC occurred at cryogenic temperatures in a grating etched on GaAs quantum wells (**FIG. 6d**), with a BIC generated at the saddle point of the band[275]. The inherited topological nature of the polariton BEC was revealed by measuring the far-field polarization vortex. In addition to the research areas discussed here, BICs have also drawn attention in fields such as topological acoustics[228,233,276,277], free-electron-light interactions[278,279], wireless power transfer[280,281], cavity optomechanics[55,282-284], and quantum photonics[84,285-287]. This will not only advance the development of photonic concepts related to BICs but also contribute to the realization of applications that rely on enhanced light–matter interactions.

**Acknowledgments**

The authors thank Z. Q. Zhang, W. Liu, J. Li and R.-Y. Zhang for discussions. M.K. and C.T.C acknowledge support from Research Grants Council (RGC) Hong Kong through grant AoE/P-502/20 and the Croucher Foundation (CAS20SC01). T.L. and M.X. acknowledge support from the National Natural Science Foundation of China (grant no. 12274332) and the National Key Research and Development Program of China (grant no. 2022YFA1404900).

**Figure captions**

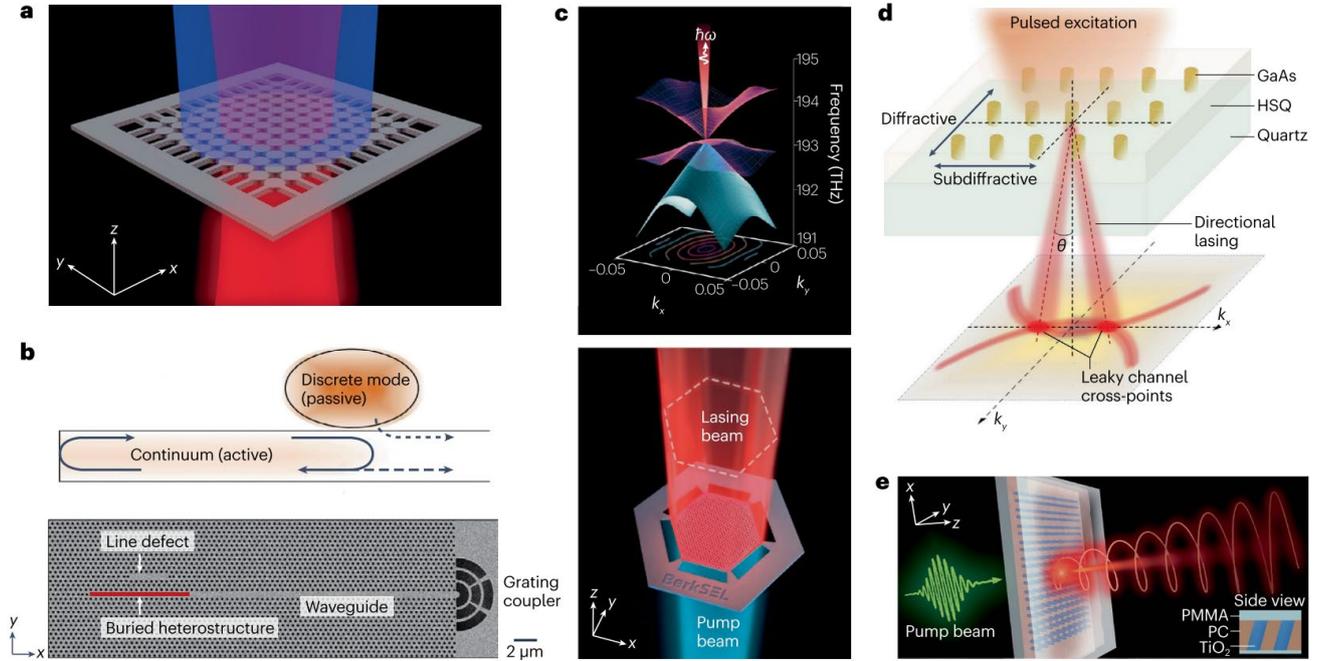

**Figure 1. Lasing boosted by the field confinement of bound states in the continuum. a** | A schematic of a bound state in the continuum (BIC) laser constructed from a suspended cylindrical nanoresonator array of multiple InGaAsP quantum wells. The pump beam (blue) incident on the array induces lasing (red). **b** | Top, the design of an ultra-coherent Fano BIC laser, which predominantly stores lasing photons in the passive region enabling it to effectively quench quantum fluctuations. The orange shading denotes the field distribution of the Fano BIC, the solid lines indicate reflection caused by the left termination of the waveguide and the side-coupled nanocavity, and the dashed lines represent destructive interference between the continuum and the discrete modes. Bottom, a scanning electron microscope image of the fabricated Fano BIC laser comprising an InP photonic crystal membrane structure with a line defect (L7 nanocavity), a buried heterostructure gain region (red), and a grating coupler at the end of the waveguide. The line defect supports discrete modes, the waveguide supports continuum modes and the grating coupler is designed to facilitate optical pumping. **c** | The design of a scalable single-mode laser. Bottom, a schematic of a photonic crystal slab (PCS) cavity that is engineered to create a zero-index BIC at a Dirac-like point. The dashed white line indicates a flat envelope of the lasing mode. Top, dispersion of the surface-emitting laser showing the B (blue) and E (pink) modes, which correspond to the irreducible representations of the $C_{6v}$ point group. The contours

in the $k_x$–$k_y$ plane ($x$ and $y$ components of the wave vector) show iso-frequency contours. The laser emission from the Dirac point with frequency $\omega$ is shown in red. The zero-index BIC enables single-mode laser emission to be preserved as the size of the cavity is scaled up. **d** | The design of a GaAs nanopillar array embedded in hydrogen silsesquioxane (HSQ) on a quartz substrate that supports directional lasing. The nanopillar array has a sub-diffractive period along the *x*-axis and a diffractive period along the *y*-axis to support diffraction orders; therefore, the symmetry-protected BIC can be turned into a leaky quasi-BIC. The directionality ($\theta$) of the laser is controlled by altering the period along the *y* direction and lasing occurs at the cross points of the two leaky channels (red curves in the emission plane $k_x$–$k_y$) under pulsed excitation of a femtosecond laser. **e** | The schematic of a metasurface that supports a chiral quasi-BIC with intrinsic chirality and giant field enhancement, leading to chiral photoluminescence and lasing (red). Inset, the chiral metasurface consists of a square lattice of tilted $TiO_2$ bars with coated films of polycarbonate (PC) and polymethyl methacrylate (PMMA) A2. Part **a** reprinted from REF.[68], Springer Nature Limited. Part **b** adapted from REF.[63], Springer Nature Limited. Part **c** adapted from REF.[100], Springer Nature Limited. Part **d** adapted from REF.[70], Springer Nature Limited. Part **e** reprinted from REF.[72], AAAS.

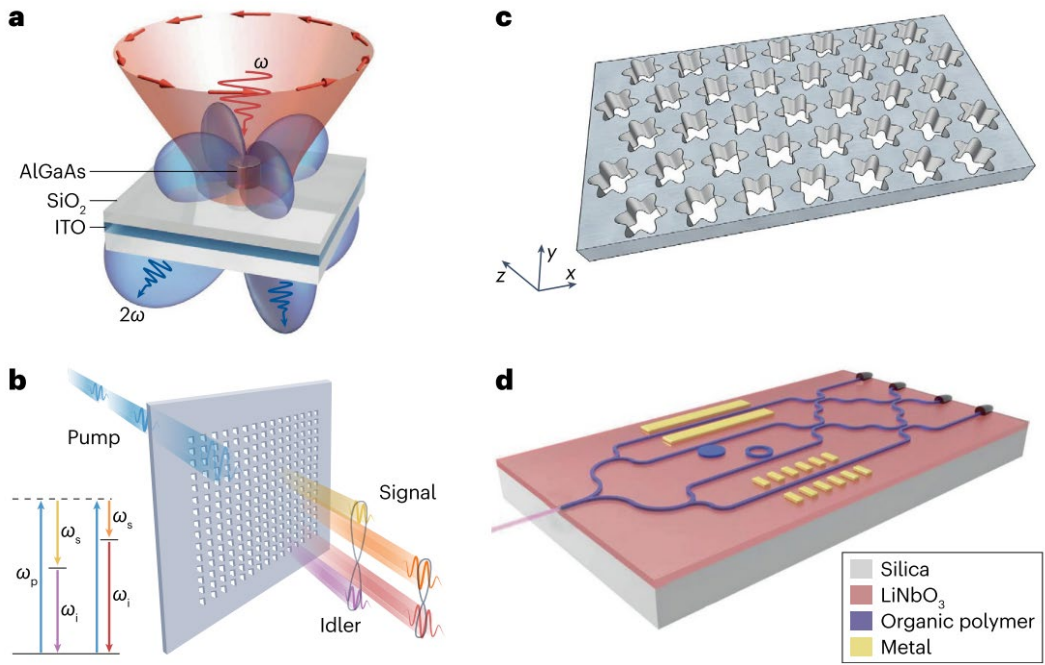

**Figure 2 Enhanced nonlinear effects and protected waveguiding with bound states in the continuum (BICs). a** | Second-harmonic generation (SHG) in a nonlinear dielectric nanoantenna pumped by an azimuthally polarized vector beam (red). The pump beam with frequency $\omega$ induces SHG with frequency $2\omega$ and a far-field radiation pattern (blue). The structure consists of a cylindrical AlGaAs nanoresonator on a three layer SiO$_2$/ITO/SiO$_2$ substrate. The dielectric resonator supports a quasi-BIC, which boosts the nonlinear conversion efficiency of SHG. **b** | Schematic of a semiconductor metasurface that supports multiple quasi-BICs and facilitates multiplexed entangled photon generation. The pump pulse with frequency $\omega_p$ produces a signal and idler pulse with frequencies $\omega_s$ and $\omega_i$, respectively, where $\omega_p = \omega_s + \omega_i$. **c** | The design for a zero-index dielectric material. The photonic crystal slab (PCS) supports three symmetry-protected BICs at a Dirac-like cone at a wave vector $k = 0$, which can substantially suppress radiation loss. **d** | An illustration of a photonic integrated circuit with BICs as waveguiding modes for eliminating energy radiation, in which low-refractive-index organic polymers are patterned on a high-refractive-index LiNbO$_3$ substrate. Part **a** reprinted from REF.[79], AAAS. Part **b** adapted from REF.[84], AAAS. Part **c** adapted with permission from REF.[85], APS. Part **d** adapted with permission from REF.[90], OSA.

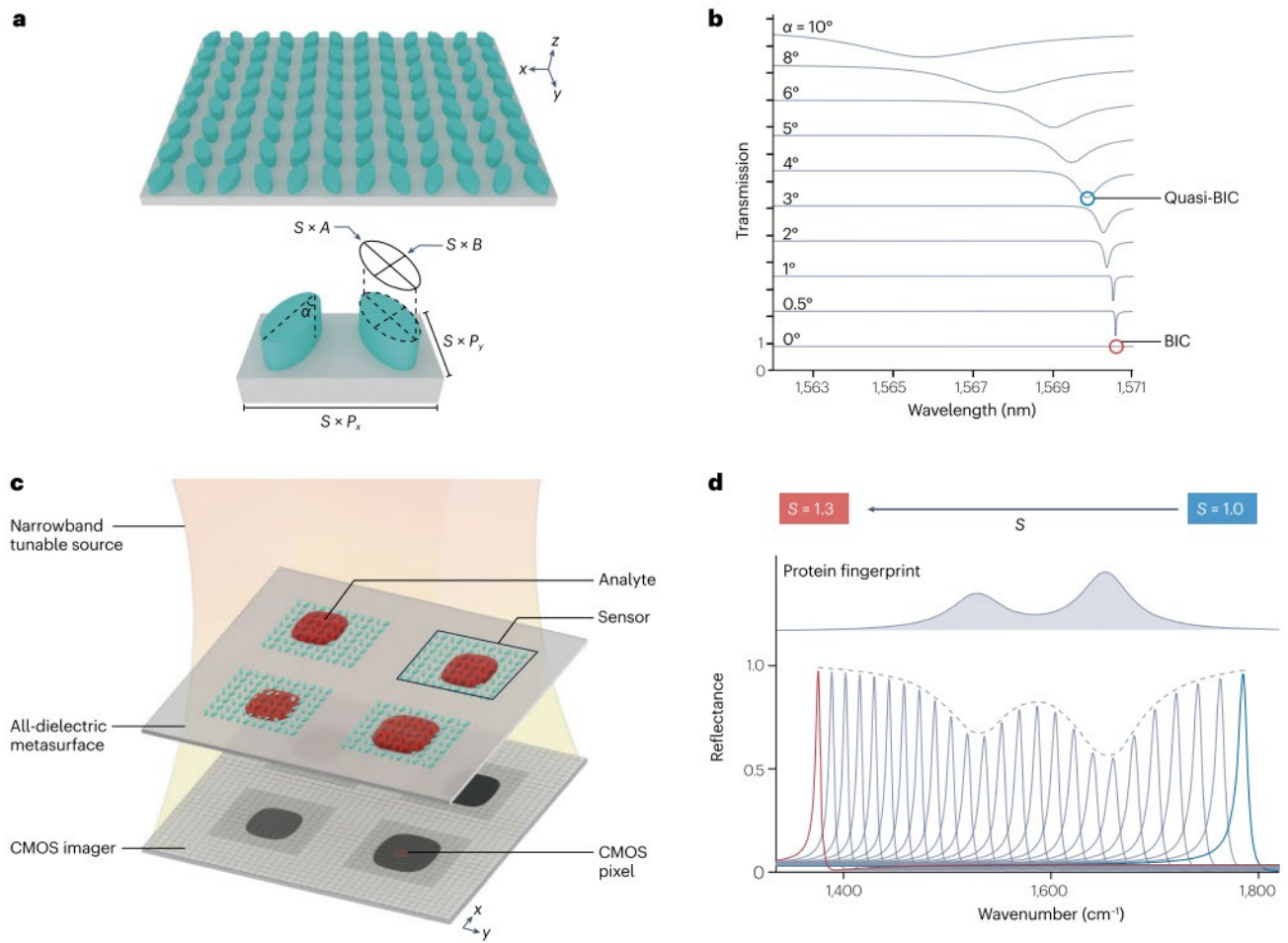

**Figure 3. The sensing benefits of Fano resonances of quasi-BICs. a** | Schematic of a metasurface comprising a square lattice of pairs of tilted silicon-bars. The symmetry-protected bound state in the continuum (BIC) turns into a quasi-BIC under the symmetry perturbation from the tilted angle $\alpha$. The resonance wavelength of the quasi-BIC is controlled by scaling the semiaxes ($A$ and $B$) and the period in the $x$ and $y$ directions ($P_x$ and $P_y$) by a factor $S$. **b** | The calculated transmission spectra of the structure in **a,** showing the evolution of the Fano lineshape at the quasi-BICs as $\alpha$ is varied between 0 and 10°. The Fano lineshape disappears at the BIC ($\alpha = 0°$). The spectra are vertically offset for clarity. **c** | A device for ultrasensitive hyperspectral imaging based on quasi-BIC metasurfaces. The sensors (cyan arrays) consist of the metasurfaces shown in **a** each with a different $S$ to tune the resonance frequency of the sensor to cover a wide range of resonance wavelengths. The system is illuminated with a narrow-band tunable laser source, and a complementary metal-oxide semiconductor (CMOS) camera records hyperspectral images at each pixel to retrieve the resonance wavelength. The resonance shifts induced by analytes can be extracted to construct spatial resonance shift maps. **d** | Simulated molecular fingerprint detection with pixelated quasi-BIC metasurfaces using the sensors shown in **c** (with $S$ being varied between 1 and 1.3). The envelope (dashed line) of the reflectance spectra (bottom) can reproduce the molecular absorption fingerprints of a protein (top). These results can be used to produce

a barcode-like absorption map. Parts **a** and **b** adapted with permission from REF.[115], APS. Part **c** adapted from REF.[139], Springer Nature Limited. Part **d** adapted from REF.[144], AAAS.

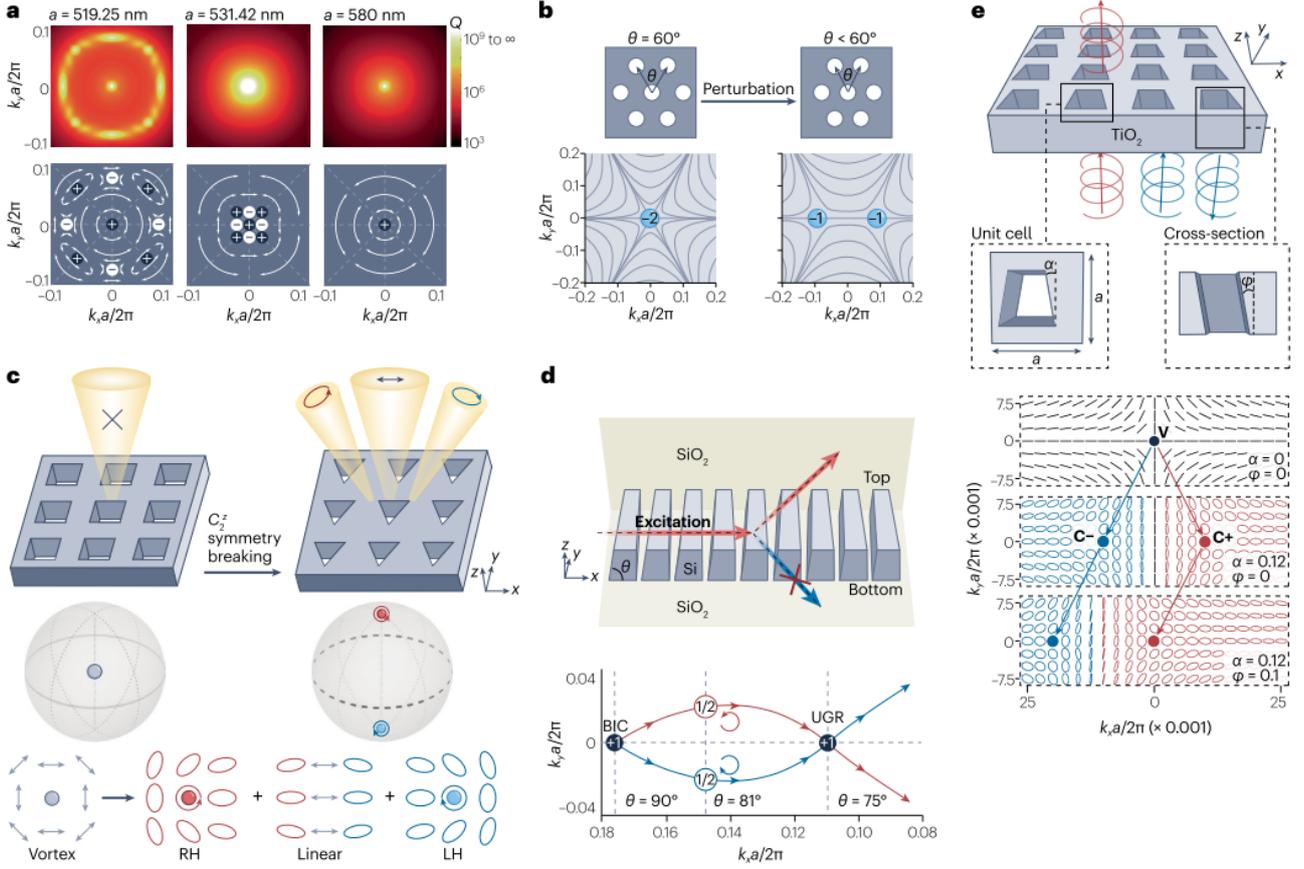

**Figure 4 Manipulating topological charges. a** | The evolution of topological charge can be used to merge bound states in the continuum (BICs). Top, the simulated $Q$ factors of a photonic crystal slab (PCS) comprising a silicon structure with a lattice of circular holes as the period $a$ of this lattice is varied. Bottom, the corresponding evolution of far-field polarization vortices (white arrows) in ($k_x$, $k_y$) momentum space. The plus and minus signs indicate vortices with a topological charge of +1 or –1, respectively. By tuning $a$ from 519.23 nm to 531.42 nm multiple BICs merge at the Γ point to form a merging BIC. By further tuning $a$ from 531.42 nm to 580 nm they annihilate in pairs and a single BIC with topological charge of +1 remains. **b** | The generation of fundamental BICs by splitting a higher-charged BIC. Top, the angle $\theta$ between the holes in the PCS is reduced from 60°, which breaks the $C_6^z$ rotation symmetry but preserves the $C_2^z$ symmetry. Bottom, this symmetry breaking splits a BIC with a topological charge of –2 (left) into two off-Γ BICs with a topological charge of –1 (right). Polarization vortices are represented by the line field tangential to the major axis of the polarization ellipse. **c** | Schematic of the generation of two circularly polarized states (C points) by breaking a BIC. Top, the in-plane inversion symmetry $C_2^z$ of a PCS with a square lattice is broken by replacing the square holes with triangular ones. The radiation (yellow cones) of the structure with a BIC is changed to exhibit elliptical and linear polarization. Bottom, this symmetry breaking destroys the BIC and creates two C points one right-handed (RH, red) and the other left-handed (LH, blue) with the same half topological

charge, which sit at opposite poles of the Poincaré sphere. There are also linearly polarized states covering the equator of the Poincaré sphere. **d** | Top, the schematic of a PCS consisting of periodic silicon bars with a sidewall angle of $\theta$. Breaking the up–down mirror symmetry ($\theta \neq 90°$) destroys the BIC; therefore, the upward (red) and downward (blue) radiation are no longer equal. Bottom, the trajectories of two C points each with a topological charge of 1/2 as $\theta$ is varied. The two C points merge at $\theta = 75°$ to form a polarization vortex on the bottom side, leading to the elimination of the downward radiation and the production of a unidirectional guided resonance (UGR). **e** | Top, schematic of a $TiO_2$ PCS with a slant perturbed unit cell defined by the angles $\alpha$ and $\phi$, which breaks both the in-plane and out-of-plane symmetries. Bottom, by varying $\alpha$ from 0 to 0.12, a BIC (V) is split into two C points one RH (C+) and the other LH (C–). By also varying $\phi$ from 0 to 0.1 a C point (C+) is tuned to the $\Gamma$ point, resulting in a quasi-BIC with intrinsic chirality. The red and blue ellipses indicate RH and LH states, the black lines indicate linear polarization. Part **a** adapted from REF.[99], Springer Nature Limited. Part **b** adapted with permission from REF.[158], APS. Part **c** adapted from REF.[107], CC BY 4.0. Part **d** adapted from REF.[108], Springer Nature Limited. Part **e** adapted from REF.[106], Springer Nature Limited.

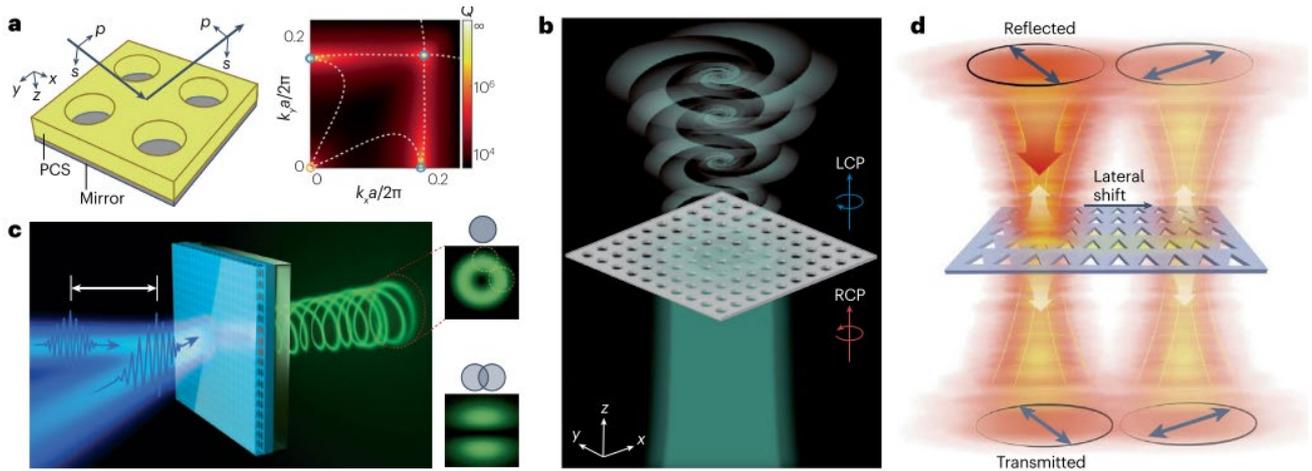

**Figure 5 Applications of polarization vortices. a** | Left, schematic of a dielectric photonic crystal slab (PCS) with an array of holes with periodicity *a* on top of a perfect mirror used to achieve complete polarization conversion. An input light with s or p polarization can be completely converted to p or s polarization, respectively, upon reflection. Right, simulated *Q* factor distribution of the guided resonances supported by the PCS in ($k_x$, $k_y$) momentum space. The dashed lines denote the critical coupling curve that defines the condition for complete polarization conversion. The yellow and blue circles indicate the symmetry-protected bound state in the continuum (BIC) and off-Γ BICs, respectively. **b** | Schematic of the optical vortex beam generated with a PCS. An incident right-handed circularly polarized (RCP) beam is used to generate a vortex beam with left-handed circular polarization (LCP) owing to the polarization vortex around the BIC. **c** | Schematic of a lead bromide perovskite metasurface with an array of circular holes on a glass substrate that is excited by two pump beams (blue) to induce vortex lasing (green) or linearly polarized lasing. Introducing a spatial deviation between the focusing spots on the surface and a time delay between the two pump pulses varies the overall intensity profile of the pumping beam making it possible to switch the output between vortex lasing and linearly polarized lasing. A spatial deviation causes asymmetry in the excitation of two pump pulses, whereas a time delay results in an ultrafast switch between symmetric excitation of a single pump pulse and asymmetric excitation of two pump pulses. The insets on the right-hand side show the far-field emission patterns that switch between a uniform ring shape (vortex lasing) and two lobes (linearly polarized lasing) under symmetric excitation and asymmetric excitation, respectively. **d** | A schematic of a PCS without in-plane inversion symmetry that induces a lateral beam shift. When a |−45 ⟩-polarized beam is directed at the PCS at normal incidence, in addition to direct reflection and transmission of the same polarized state, the converted orthogonal polarized states acquire a geometric phase gradient in momentum space and are therefore laterally shifted relative to the incident beam in the direction opposite to the phase gradient. Part **a** adapted with permission from REF.[184], APS. Part **b** adapted

from REF.[166], Springer Nature Limited. Part **c** adapted from REF.[168], AAAS. Part **d** adapted from REF.[171], Springer Nature Limited.

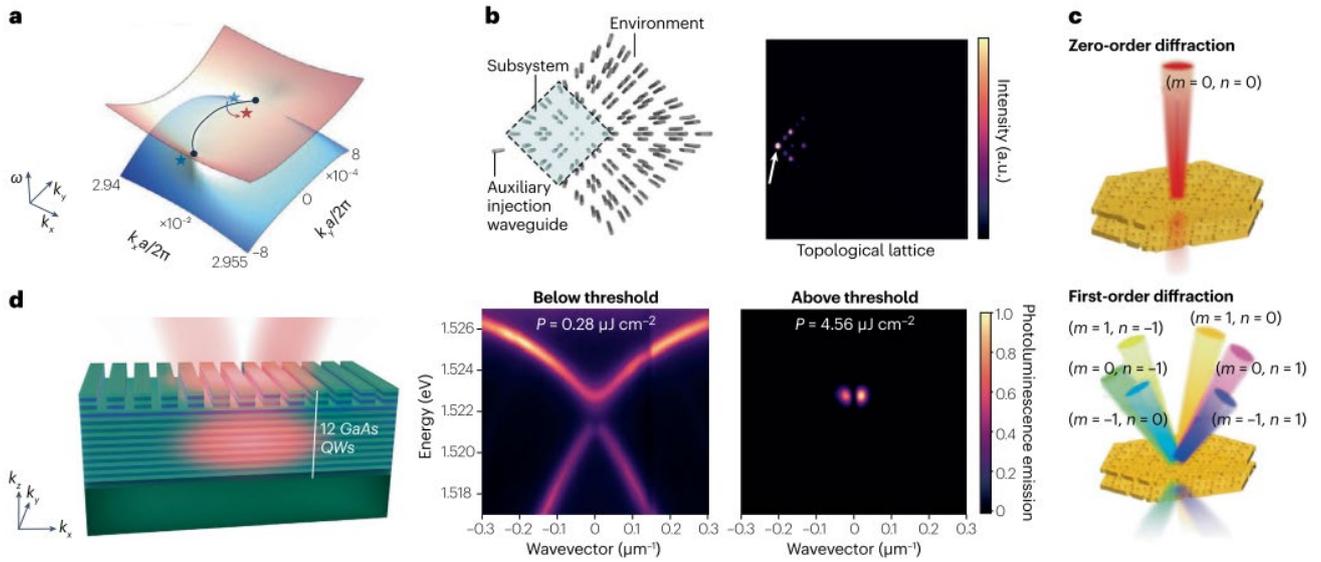

**Figure 6 Emerging research areas. a** | The interplay between circularly polarized states (C points, stars) and a bulk Fermi arc (black line) emerging from paired exceptional points (black dots). A C point can be tuned to cross the bulk Fermi arc from the lower band to the upper band on simulated band structures to modulate the topological charge of the bulk Fermi arc by varying the structural parameters. The blue and red stars indicate C points on the lower and upper band, respectively. ω indicates frequency, and $k_x$ and $k_y$ are $x$ and $y$ components of a wave vector, respectively. **b** | Left, schematic of a second-order topological insulator waveguide array. Right, experimentally observed light intensity when light is injected into the system through the auxiliary waveguide (white arrow). The topological corner state remains localized in the subsystem (dashed box on the left) and does not couple with the bulk modes supported by the environment. **c** | Quasi-BICs supported by moiré superlattices on a twisted bilayer photonic crystal slab (PCS). Symmetry-protected BICs are present at zero-order diffraction (top), whereas first-order diffraction (bottom) and other existing higher-order diffraction modes are radiative. ($m$, $n$) indicates the reciprocal lattice vector. **d** | Left, schematic of a multilayer planar waveguide with a grating etched on 12 GaAs quantum wells (QWs). Right, the measured angle-resolved photoluminescence emission under non-resonant pulsed excitation. Below the threshold for polariton Bose–Einstein condensation the exciton–polariton dispersion has a polariton BIC (dark spot) at 0 μm$^{-1}$ on the lower branch. Above the threshold the exciton–polariton dispersion becomes dominated by a double-peaked emission, in which the condensate emission comes from the polariton BIC. Part **a** adapted with permission from REF.[217],PNAS. Part **b** adapted with permission from REF.[230], APS. Part **c** adapted with permission from REF.[246], APS. Part **d** adapted from REF.[275], Springer Nature Limited.

**Box 1 Topological features**

**Topological charge**

The topological concept of bound states in the continuum (BICs) can be illustrated using a photonic crystal slab[8] (PCS), in which the far-field radiation is described by a two-component vector $\mathbf{E}_{//} = (E_x, E_y)$ with a wavector $\mathbf{k} = (k_{//}, k_z)$ (panel **a**). Here $E_x$ and $E_y$ are the projection of the radiating plane wave $\mathbf{E}$ along the *x* and *y* directions, respectively. $k_{//}$ is the projection of the wavevector along the *x-y* plane and has two components $k_x$ and $k_y$. In a 2D momentum space spanned by $k_x$ and $k_y$, $E_x = 0$ and $E_y = 0$ each correspond to a continuous curve and the intersection of these two curves represents the vanishing of far-field radiation, that is, a BIC (panel **b**), which is at the center of the polarization vortices of $\mathbf{E}_{//}$. To characterize the topology of this singular point the topological charge or winding number of $\mathbf{E}_{//}$ is calculated along a closed anticlockwise loop surrounding the BIC (black circle in panel **b**). When the winding number is –2 the vector field direction with an angle $\theta$ rotates clockwise twice each time it travels around the loop (panel **c**). There are generally several vanishing points for $\mathbf{E}_{//}$ in the $(k_x, k_y)$ space. The winding number along any closed loop is always conserved during continuous parameter variation unless that loop passes a singular point, such as, a BIC.

**Polarization vortices**

When the phase between $E_x$ and $E_y$ can be freely tuned by breaking the $C_2^z$ symmetry, far-field radiation becomes elliptically polarized and the winding number defined above with linearly polarized states is not appropriate. Conventionally, the north and south poles of the Poincaré sphere denote circularly polarized states, the equator represents linearly polarized states, and the hemispheres represent elliptically polarized states. The polarization states can be described by the Stokes parameters $S_1$, $S_2$ and $S_3$. The additional phase freedom lifts the polarization states from the equator to an arbitrary location on the sphere. Hence a nontrivial winding along the equator can be continuously shrunk to a point (panel **d**). Alternatively, the long axis direction of the elliptical polarization can be used to define the charge of the right and left circularly polarized states (panels **e** and **f**), where the polarization singularity has a half charge.

**Geometric phase**

When a polarization state undergoes a series of polarization conversions then returns to its initial polarization, it traces a closed loop on the Poincaré sphere and gains a Pancharatnam–Berry phase. The solid angle enclosed by this loop is equal to twice the geometric phase. A polarization state can be converted from right to left circular polarization through the scattering of guided resonances in a PCS with linear far-field polarization. This process corresponds to a path connecting the north pole, the linear polarization on the equator, and the south pole of a Poincaré sphere (panel **g, up**). The path changes accordingly as guided resonances exhibit different far-field polarization direction $\theta$. The solid angle $\Omega$ subtended by two paths is equivalent to twice their geometric phase difference. When a polarization state is converted from left to right circular polarization, the path has an opposite direction and the sign of the geometric phase changes (panel **g, down**).

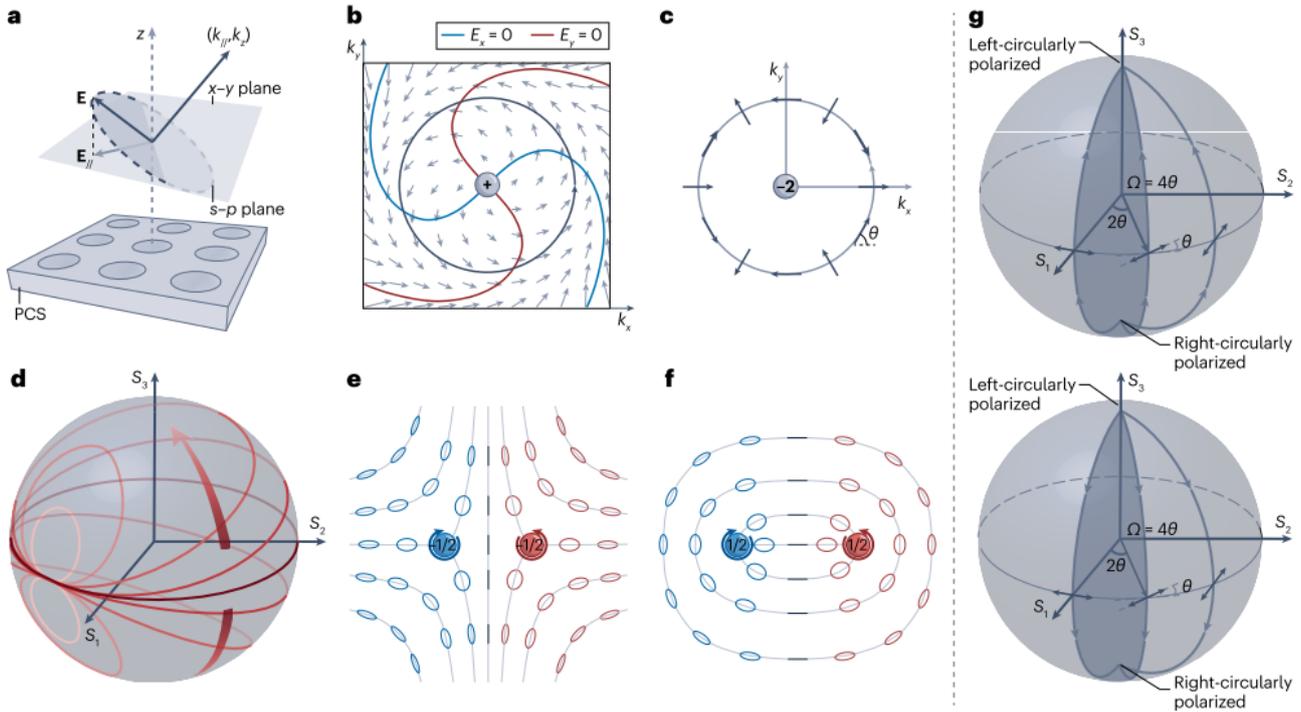

**Glossary**

**Photonic crystal slab**
A periodic planar waveguide that possesses 2D periodicity but has a finite thickness.
**Zero-index material**
Materials that have zero refractive index at certain frequencies.
$C_{4v}$ **point group**
A symmetry group with rotations and reflections associated with a square.
**Mie resonance**
Resonances that occur in the scattering of particles with sizes comparable to the wavelength of the incident light.
**Fabry–Pérot resonance**
Resonances supported by two parallel reflecting surfaces.
**Nonlinear frequency conversion**
A type of nonlinear optical effect, which includes processes such as second harmonic generation, third harmonic generation, high harmonic generation, sum-frequency generation, difference-frequency generation and spontaneous parametric down-conversion.

**Fano interference**

Interference between discrete bound states and a continuum of extended states, resulting in the generation of Fano resonance.

**$C_{6v}$ point group**

A symmetry group with rotations and reflections associated with a hexagon.

**Goos–Hänchen shift**

A lateral shift of the reflected and transmitted beams in the direction parallel to the incident plane

**Imbert–Fedorov shift**

A spin-dependent lateral shift of the reflected and transmitted beams in the direction perpendicular to the incident plane.

**Bulk Fermi arc**

In classical systems, a bulk Fermi arc refers to an isofrequency contour of bulk band structures that presents as an arc in momentum space.